\documentclass[aps,twocolumn,nofootinbib,showpacs]{revtex4}

\def\Lie{\pounds}
\def\dual{{}^\star\!}

\begin{document}

\preprint{gr-qc/0508026}

\title{The Hamiltonian boundary term and quasi-local energy
flux
}

\author{Chiang-Mei Chen} \email{cmchen@phy.ncu.edu.tw}
\affiliation{Department of Physics, National Central University,
Chungli 32054, Taiwan}

\author{James M. Nester} \email{nester@phy.ncu.edu.tw}
\affiliation{Department of Physics and Institute of Astronomy,
National Central University, Chungli 32054, Taiwan}

\author{Roh-Suan Tung}\email{tung@shnu.edu.cn}
\affiliation{Center for Astrophysics, Shanghai Normal University,
100 Guilin Road, Shanghai 200234, China}

\date{8 August 2005}


\begin{abstract}
The Hamiltonian for a gravitating region includes a boundary term
which determines not only the quasi-local values but also, via the
boundary variation principle, the boundary conditions. Using our
covariant Hamiltonian formalism, we found four particular
quasi-local energy-momentum boundary term expressions; each
corresponds to a physically distinct and geometrically clear
boundary condition. Here, from a consideration of the asymptotics,
we show how a fundamental Hamiltonian identity naturally leads to
the associated quasi-local energy flux expressions. For
electromagnetism one of the four is distinguished:  the only one
which is gauge invariant; it gives the familiar energy density and
Poynting flux. For Einstein's general relativity two different
boundary condition choices correspond to quasi-local expressions
which asymptotically give the ADM energy, the Trautman-Bondi
energy and, moreover, an associated energy flux (both outgoing and
incoming). Again there is a distinguished expression: the one
which is covariant.

\end{abstract}


\maketitle


\section{Introduction}

The Hamiltonian that generates the dynamical evolution of any
physical system within a (finite or infinite) spatial region
necessarily includes, in addition to an integral over the spatial
volume, an integral over the boundary of the region. Our concern
is with this Hamiltonian boundary term: we wish to understand both
its role and proper form.

The Hamiltonian is, of course, related to energy.  Every physical
system carries mass-energy and consequently inevitably generates
gravity, hence gravity plays the major role in our analysis. We
consider only geometric gravity theories, and here more
specifically only Einstein's general theory of relativity (GR).

The essential necessity for, and the role of, the Hamiltonian
boundary term for asymptotically flat spaces in GR was first
clearly discussed in a seminal work of Regge and Teitelboim
\cite{RT74}. They argued that in order for the Hamiltonian to be
functionally differentiable on the phase space of asymptotically
flat spatial metrics, certain boundary integrals over the 2-sphere
at spatial infinity must be included.  These turned out to be in
fact just the ADM \cite{ADM} asymptotic expressions for the
conserved total quantities: energy, momentum and angular momentum
(plus an additional expression for the center-of-mass).
Subsequently certain improvements were made in the analysis by
Beig and \'O Murchadha \cite{BO87}; more recently Szabados has
made some further refinements \cite{Sza03}.

Although the total conserved quantities are well defined for
asymptotically flat spaces, as is well known, there are no well
defined local densities for these quantities. This can be
understood in terms of the equivalence principle, which precludes
the detection of gravity at a point (see, e.g. \cite{MTW} \S
20.4). The modern idea is that one should have quasi-local
quantities: i.e., quantities associated with a closed 2-surface
(for a nice review of this topic see \cite{Sza04}).

For finite regions, we found that the value of the Hamiltonian
boundary term can determine the quasi-local quantities. Using
essentially the same principle as used by Regge-Teitelboim (this
principle will play the major role in our subsequent discussion)
we found that the Hamiltonian boundary term also determines the
boundary conditions. Using this idea along with our covariant
Hamiltonian formalism, for each dynamic field we found four types
of quasi-local energy-momentum boundary term expressions; each
corresponds to a physically distinct and geometrically clear
boundary condition \cite{Chen:1994qg,Chen:1998aw,Chen:2000xw}.

However a consideration of the phase space asymptotics necessary
for a well defined Hamiltonian naturally leads to the recognition
that the conditions {\it are not met} in the radiating regime.
This apparent difficulty provides the opportunity for our main new
result. We show here how a fundamental Hamiltonian identity
naturally leads, for each Hamiltonian boundary expression, to an
associated quasi-local energy flux expression.

We note that gravitational energy flux (aka balance) expressions
have a long history, going back to Trautman \cite{Trautman}, Bondi
\cite{Bondi} and Sachs \cite{Sachs}.  Nevertheless there still is
considerable interest in this topic and some significant progress
has been made recently; see in particular
\cite{Hay94,Katz97,BLY97,CJM98,CJK02,MFC03,Yoon04}. We include
here some applications of our new natural energy flux expressions.

An application of our formalism with its quasi-local energy and
energy flux expressions to the electromagnetic field, where we can
readily interpret the expressions and boundary conditions, is
included to illustrate the ideas in a familiar setting. For the
electromagnetic case one of our four expressions is distinguished:
the only one which is gauge invariant; it gives the usual energy
density and Poynting flux.

When applied to Einstein's general relativity two different
boundary condition choices correspond to quasi-local expressions
which asymptotically give the ADM energy, the Trautman-Bondi
energy and, moreover, an associated energy flux (both outgoing and
incoming). However once again there is a distinguished expression:
in this case it is the one which is covariant.

The plan of this work is as follows: In section II we discuss the
first order Lagrangian formalism. In section III we consider local
translational invariance. The basic Hamiltonian formulation is
considered in section IV.  In section V we consider  refined
boundary terms. There follows a discussion of the phase space and
the asymptotics in Section VI.   Our new flux expressions are
presented in section VII.  In section VIII we consider the
application of these ideas to electromagnetism. The application to
Einstein's gravity theory is covered in section IX.  A discussion
forms the concluding section.

\section{The first order Lagrangian formalism}
Our formalism is intended to be applicable to general types of
fields. Technically we find it convenient to represent dynamic
fields in terms of differential forms.  (More precisely to
accommodate all the fields found in nature we would use a
collection of tensor and spinor valued forms including Dirac
spinors and gauge potential one-forms along with the spacetime
orthonormal coframe and connection one-forms).   Here we develop
the representative case of an k-form field $\varphi$ (the field
may take its values in some vector space and may thus carry some
indices which are here suppressed; the generalization to include
several fields, possibly of different grades, is straightforward).

We proceed from the action principle.  It can be shown that any
action principle can be rewritten in an equivalent form, which
(following e.g.~\cite{ADM,Kuc76}) we refer to as {\it first
order}; this is the most convenient form for our purposes. A {\it
first order Lagrangian 4-form} for a k-form field $\varphi$ has
the form
\begin{equation}
{\cal L} = d \varphi \wedge p - \Lambda(\varphi,p).\label{1ordL}
\end{equation}
Its variation has the form
\begin{equation} \label{varL}
\delta {\cal L} = d(\delta \varphi \wedge p) + \delta \varphi
\wedge \frac{\delta {\cal L}}{\delta \varphi} + \frac{\delta {\cal
L}}{\delta p} \wedge \delta p,
\end{equation}
implicitly defining {\it a pair of first order} Euler-Lagrange
expressions, which are explicitly given by
\begin{equation}
\frac{\delta {\cal L}}{\delta p} := d \varphi - \partial_p
\Lambda, \qquad \frac{\delta {\cal L}}{\delta \varphi} := -
\varsigma d p - \partial_\varphi \Lambda, \label{1ordfe}
\end{equation}
where $\varsigma:=(-1)^k$.  The integral of ${\cal L}$ associates
an action with any spacetime region.  The variation of this action
is given by the integral of $\delta{\cal L}$. The total
differential term in (\ref{varL}) then leads to an integral over
the boundary of the region. Hamilton's principle---that the action
should be extreme with $\delta\varphi$ vanishing on the
boundary---yields the field equations: the vanishing of the
Euler-Lagrange expressions (\ref{1ordfe}).

\section{Local translation invariance}  The action should not
depend on the particular way points are labeled.  Thus it should
be invariant under diffeomorphisms, including infinitesimal
diffeomorphisms---which correspond to a displacement along some
vector field $N$.  From a gauge theory perspective such
displacements are regarded as a ``local translation''. Under a
local translation quantities change according to the Lie
derivative. Hence, for a diffeomorphism invariant action the
relation (\ref{varL}) should be identically satisfied when the
variation operator $\delta$ is replaced by the Lie derivative
$\Lie_N$ ($\equiv di_N+i_Nd$ on the components of form fields):
\begin{equation}
d i_N {\cal L} \equiv \Lie_N{\cal L} \equiv  d(\Lie_N \varphi
\wedge p) + \Lie_N \varphi \wedge \frac{\delta {\cal L}}{\delta
\varphi} + \frac{\delta {\cal L}}{\delta p} \wedge \Lie_N
p.\label{noethertrans}
\end{equation}
This simply means that ${\cal L}$ is a 4-form which depends on
position only through the fields $\varphi$, $p$. (For this to be
the case the set of fields in ${\cal L}$ necessarily includes
dynamic geometric variables, which means gravity.)

 From (\ref{noethertrans}) it directly follows that the 3-form
\begin{equation}
{\cal H}(N) := \Lie_N \varphi \wedge p - i_N {\cal L} \label{HN}
\end{equation}
satisfies the identity
\begin{equation} \label{IdH}
- d {\cal H}(N) \equiv  \Lie_N \varphi \wedge \frac{\delta {\cal
L}}{\delta \varphi} + \frac{\delta {\cal L}}{\delta p} \wedge
\Lie_N p,
\end{equation}
showing that it is a conserved ``current'' {\it on shell}
(meaning: when the field equations are satisfied). Substituting
(\ref{1ordL}) into (\ref{HN}) leads to the explicit expression $
{\cal H}(N)\equiv d(i_N \varphi \wedge p) + \varsigma i_N \varphi
\wedge d p + \varsigma d \varphi \wedge i_N p + i_N \Lambda $,
 from which one can see that this conserved  {\it Noether
translation current} can be written as a 3-form linear in the
displacement vector plus a total differential:
\begin{equation}
{\cal H}(N) =: N^\mu {\cal H}_\mu + d {\cal B}(N). \label{H+dB}
\end{equation}
Compare the differential of this expression, $ d{\cal H}(N) \equiv
dN^\mu\wedge {\cal H}_\mu + N^\mu d {\cal H}_\mu, $ with
(\ref{IdH}); equating the $dN^\mu$ coefficient on both sides
reveals that
\begin{equation}
{\cal H}_\mu \equiv - i_\mu \varphi \wedge \frac{\delta {\cal
L}}{\delta \varphi} + \varsigma \frac{\delta {\cal L}}{\delta p}
\wedge i_\mu p.
\end{equation}
This identity is a necessary consequence of {\it local}
diffeomorphism invariance (i.e., symmetry for non-constant
$N^\mu$). From this relation one can see that ${\cal H}_\mu$
vanishes on shell; hence the value of the conserved quantity
associated with a local displacement $N$ and a spatial region
$\Sigma$ is determined by a 2-surface integral over the boundary
of the region:
\begin{equation}
E(N,\Sigma) := \int_\Sigma {\cal H}(N) = \oint_{\partial\Sigma}
{\cal B}(N). \label{EN}
\end{equation}
The value is {\em quasi-local}.  For {\it any} choice of $N$ this
expression defines a conserved quasi-local quantity.

What do these values mean? In general we do not have a clear
physical interpretation. However, at least if the geometry on the
boundary is not so far from flat space, for a suitable
timelike(spacelike) quasi-translation displacement on the boundary
the expression defines a quasi-local energy(momentum), and for a
suitable quasi-rotation(boost) it defines a quasi-local angular
momentum(center-of-mass).  (Here we do not explore the important
question of how to specifically make these quasi-displacement
choices for a general region in order to obtain good physical
quasi-local values.)

\section{The Hamiltonian Formulation}

 From the first order field equations (\ref{1ordfe}), by
contraction with a ``time evolution vector field'' $N$ and using
$i_Nd\varphi=\Lie_N\varphi-di_N \varphi$, $i_Ndp=\Lie_N p-di_N p$,
we get a pair of Hamiltonian-like evolution equations for the
``time derivatives'': $\Lie_N\varphi$, $\Lie_N p$. A key identity
involving these time derivatives is revealed by comparing two
relations; on the one hand take the projection of $\delta L$
(\ref{varL}) along $N$:
\begin{eqnarray}
i_N\delta{\cal L}&\equiv& i_N d(\delta \varphi \wedge p) +
i_N\left[\delta \varphi \wedge \frac{\delta {\cal L}}{\delta
\varphi} + \frac{\delta {\cal
L}}{\delta p} \wedge \delta p\right]
\nonumber\\
&\equiv& \Lie_N(\delta\varphi\wedge p)- di_N(\delta \varphi \wedge
p)
\nonumber\\
&& + i_N\left[\delta \varphi \wedge \frac{\delta {\cal L}}{\delta
\varphi} + \frac{\delta {\cal L}}{\delta p} \wedge \delta p\right]
;
\end{eqnarray}
on the other hand take the projection of the Lagrangian 4-form
$i_N{\cal L}$, which from (\ref{HN}) is just $\Lie_N\varphi\wedge
p-{\cal H}(N)$, and vary:
\begin{eqnarray}
\delta i_N {\cal L} &\! \equiv\! & \delta (\Lie_N \varphi \wedge
p) - \delta{\cal H}(N)
\\
&\! \equiv\! & \delta (\Lie_N\varphi) \wedge p + \Lie_N \varphi
\wedge \delta p - \delta{\cal H}(N)
\nonumber \\
&\! \equiv\! & \Lie_N \delta \varphi \wedge p + \Lie_N \varphi
\wedge \delta p - \delta{\cal H}(N)
\nonumber \\
&\! \equiv\! & \Lie_N(\delta\varphi \wedge p) - \delta \varphi
\wedge \Lie_N p + \Lie_N \varphi \wedge \delta p - \delta{\cal
H}(N). \nonumber
\end{eqnarray}
Since $N$ is not varied the two relations are identical: $\delta
i_N{\cal L}\equiv i_N\delta{\cal L}$; consequently,
\begin{eqnarray}
\delta {\cal H}(N) &\equiv& - \delta \varphi \wedge \Lie_N p +
\Lie_N \varphi \wedge \delta p
\\
&& + d i_N(\delta \varphi \wedge p) - i_N \left[\delta \varphi
\wedge \frac{\delta {\cal L}}{\delta \varphi} + \frac{\delta {\cal
L}}{\delta p} \wedge \delta p\right]. \label{varH} \nonumber
\end{eqnarray}
The last term vanishes ``on shell''. This relation identifies the
Noether translational current ${\cal H}(N)$ as the {\em
Hamiltonian 3-form} (i.e., density), as the following
considerations show. The integral of ${\cal H}(N)$ over a
3-dimensional region,
\begin{equation}
H(N,\Sigma):=\int_\Sigma {\cal H}(N), \end{equation} is the
Hamiltonian which displaces this region along $N$, since the
integral of its variation:
\begin{equation}
 \delta
H(N,\Sigma)=\int_\Sigma \delta {\cal H}(N),
\end{equation}
yields, from (\ref{varH}), (on shell) the Hamilton equations:
\begin{equation}
\Lie_N\varphi=\frac{\delta{\cal H}(N)}{\delta p},\qquad \Lie_N
p=-\frac{\delta{\cal H}(N)}{\delta \varphi}, \label{hameqs}
\end{equation}
{\em if} the boundary term in the variation of the Hamiltonian
{\em vanishes}. In this case that means when $\delta\varphi$
vanishes on $\partial\Sigma$. Technically the variational
derivatives of the Hamiltonian $H(N,\Sigma)$ are only defined for
variations satisfying this boundary condition.  In other words
this Hamiltonian is functionally differentiable on the phase space
of fields satisfying this particular boundary condition.

\section{Refined boundary terms}

In some important cases the fields of physical interest do not
satisfy the  aforementioned boundary  condition naturally
inherited from the Lagrangian (the main example is the spacetime
metric for an asymptotically flat region). A suitably modified
formulation is needed to deal with this situation.  One
alternative is to modify the Lagrangian 4-form itself by a total
differential.  As discussed in some detail in \cite{Chen:2000xw},
this would modify the boundary condition on the whole
3-dimensional boundary of the spacetime region, thus inducing the
same type of modification on the spatial boundary at large spatial
distances as on the initial time hypersurface.  However we
actually want the freedom to adjust the boundary condition on the
2-dimensional boundary of the spacelike region $\partial\Sigma$
independently of the type of initial conditions imposed within
$\Sigma$. For this purpose we focus on the Hamiltonian. We note
that the Hamiltonian (\ref{H+dB}) has two distinct parts; each
plays a distinct role. The proper density $N^\mu{\cal H}_\mu$,
although it has vanishing value on shell, generates the equations
of motion, whereas the boundary term ${\cal B}(N)$ determines both
the {\em quasi-local value} (\ref{EN}) and the {\em boundary
condition}. Now the boundary term can be adjusted--- without
changing the Hamilton equations or the conservation property
(\ref{IdH})---indeed we can replace the 2-form ${\cal
B}(N)=i_N\varphi\wedge p$ inherited from the Lagrangian by any
other. Such an adjustment is in one respect just a special case of
the conserved Noether current ambiguity (i.e. for any 2-form
$\chi$, $J$ and $J':=J+d\chi$ are both conserved currents
(3-forms) if $dJ=0$, even though they define different conserved
values). However here, in this Hamiltonian case, any such
adjustment modifies---{\em in parallel}---not only the value of
the quasi-local quantities but also the spatial boundary
conditions. Thus the boundary term ambiguity is under physical
control: each distinct choice of Hamiltonian boundary quasi-local
expression is associated with a physically distinct boundary
condition \cite{Chang:1998wj,Chen:1998aw,Chen:2000xw,Nes04}. We
thus find that the Hamiltonian density always takes the general
form
\begin{equation}
{\cal H}_{\rm B.C.}(N) = N^\mu {\cal H}_\mu + d {\cal B}_{\rm
B.C.}(N) \, \label{Hbc}.
\end{equation}
The subscript ``B.C.'' here refers to the particular choice of
built in boundary condition.

In order to accommodate suitable boundary conditions we found
that, in general, one must introduce (at the minimum actually only
on the boundary, but to simplify the discussion here we presume it
to be on any desired neighborhood of the boundary) certain
reference values $\bar p$, $\bar \varphi$, which represent the
ground state of the field---the ``vacuum'' (or background field)
values (this is necessary in particular for fields whose natural
ground state is non-vanishing, e.g.\ the spacetime metric).  We
take our boundary terms to be linear in
$\Delta\varphi:=\varphi-\bar\varphi$ and $\Delta p:=p-\bar p$, so
that they (and thus all the quasi-local quantities) vanish if the
fields take on the ground state (i.e.~reference) values. We
presume that the reference values (like $N$) are not varied:
$\delta\bar\varphi=0$ and $\delta\bar p=0$, consequently
$\delta\Delta\varphi=\delta\varphi$, $\delta\Delta p=\delta p$.
Note that we presume the reference values (as well as $N$) to be
defined on the dynamic spacetime in the region of interest,
independently of the dynamic fields; they are regarded as being
fixed prior to (and thus independently of) the choice of $\Sigma$
and $S=\partial\Sigma$.

 From our investigations \cite{Chen:1994qg,Chen:1998aw} we found
two ``covariant'' boundary term alternates to ${\cal
B}(N):=i_N\varphi\wedge p$, namely
\begin{eqnarray}
{\cal B}_{\rm Dirichlet}(N) &:=& i_N \varphi \wedge \Delta p -
\varsigma \Delta \varphi \wedge i_N \bar p, \label{Bdir}\\
{\cal B}_{\rm Neumann}(N) &:=& i_N \bar \varphi \wedge \Delta p -
\varsigma \Delta \varphi \wedge i_N p.\label{Bneu}
\end{eqnarray}
A short calculation (of the form $\delta {\cal H}_{\rm
B.C.}=\delta ({\cal H}-d{\cal B})+\delta d{\cal B}_{\rm B.C.}$)
shows that the variation of the associated Hamiltonian 3-forms
(\ref{Hbc}) have (on shell) the respective forms
\begin{eqnarray}
\delta{\cal H}_{\rm Dirichlet}(N) &\equiv& - \delta \varphi \wedge
\Lie_N p + \Lie_N \varphi \wedge \delta p
\nonumber\\
&& + di_N(\delta \varphi\wedge \Delta p), \label{deltaHdir}
\\
\delta{\cal H}_{\rm Neumann}(N)&\equiv& - \delta \varphi \wedge
\Lie_N p + \Lie_N \varphi \wedge \delta p
\nonumber\\
&& - di_N(\Delta \varphi\wedge \delta p)\label{deltaHneu},
\end{eqnarray}
corresponding to holding fixed certain (after integration certain
projected) covariant sets of components, respectively the value of
the field and the value of its canonically conjugate momentum;
whence our labels.

Moreover, we found two more physical interesting choices
\cite{Chen:2000xw}
\begin{eqnarray}
{\cal B}_{\rm dynamic}(N) &:=& i_N \bar\varphi \wedge \Delta p -
\varsigma \Delta \varphi \wedge i_N \bar p, \label{Bdyn}\\
{\cal B}_{\rm constrained}(N) &:=& i_N \varphi \wedge \Delta p -
\varsigma \Delta \varphi \wedge i_N p.\label{Bcon}
\end{eqnarray}
The variation of the associated Hamiltonian 3-forms have (on
shell) the indicated  forms
\begin{eqnarray}
\delta{\cal H}_{\rm dynamic}(N) &\equiv& - \delta \varphi \wedge
\Lie_N p + \Lie_N \varphi \wedge \delta p \\
&& + d (\varsigma \delta \varphi \wedge i_N \Delta p - i_N \Delta
\varphi \wedge \delta p), \label{deltaHdyn} \nonumber
\\
\delta{\cal H}_{\rm constrained}(N)&\equiv&  - \delta \varphi
\wedge \Lie_N p + \Lie_N \varphi \wedge \delta p
\\
&& + d ( \delta i_N\varphi \wedge \Delta p - \varsigma \Delta
\varphi \wedge \delta i_N p)\label{deltaHcon}, \nonumber
\end{eqnarray}
corresponding to holding fixed certain (after integration certain
projected) components of respectively the ``dynamic parts'' (the
spatial pullback) of $\varphi$, $p$ and the ``constrained parts''
(the ``time'' projections) $i_N\varphi$, $i_N p$.

In each of these cases the boundary term in the Hamiltonian
variation has a certain {\it symplectic} structure which pairs
certain {\em control} and {\em response} quantities \cite{KT79}.
Within each of our two sets of expressions the pairs are simply
related by an interchange of ``control'' and ``response'',
formally $\delta\to\Delta$, $\Delta\to-\delta$.

Note that one expression stands out: for fields which allow
trivial reference values, $\bar \varphi=0=\bar p$, the boundary
term ${\cal B}_{\rm dynamic}(N)$ vanishes.  These fields, with
this choice of boundary condition, make no explicit contribution
to the quasi-local boundary term. Thus there is a certain
preferred boundary expression---and thus {\em a preferred boundary
condition}---for this large class of fields, a class which
includes all the necessary physical fields---aside from dynamic
geometry gravity.

An instructive discussion of boundary conditions associated with
variational principles appears in Lanczos \cite{Lanczos}, section
II.15. The variational principle always gives us the correct
number of boundary conditions.  However it should be remarked that
it cannot {\it guarantee} that the boundary conditions are the
proper ones for the existence and uniqueness of solutions; that
would depend on the particular quality and type of the equations,
which is influenced especially by the metric signature (our
formalism ``formally'' does not care) and the details of the
Lagrangian ``potential'' $\Lambda(\varphi,p)$. Our general
formalism cannot take such particulars into account.

Our philosophy is that normally one should ``control'' on the
boundary the indicated variations. Lanczos also discusses an
interesting alternative: one could instead take ``free''
variations on the boundary.  To have the boundary term in the
variation vanish with free variations of the ``control'' variables
one must then require that the associated ``response'' vanishes;
this yields what is referred to as the {\em natural boundary
condition}. For our expressions it would simply mean that the
response fields would take, on the boundary, the pre-specified
``reference'' values. Hence, as far as imposing boundary
conditions is concerned, one can achieve the same result by these
two different approaches: (i) control the variable to the desired
boundary value, or (ii) take free variations with the appropriate
reference fields chosen to have the desired boundary values. Note,
however, that there are differences in the resultant quasi-local
values.  In particular a free variation for ${\cal H}_{\rm
Dirichlet}$ requires from (\ref{deltaHdir}) that $p=\bar p$.
Consequently the value of the Hamiltonian is determined by a
reduced form of the boundary expression (\ref{Bdir}): $-\varsigma
\Delta\varphi\wedge i_Np$. Similarly, free variation for ${\cal
H}_{\rm Neumann}$ leads to $\varphi=\bar\varphi$ and the reduced
boundary term $i_N\varphi\wedge\Delta p$.  On the other hand, for
${\cal H}_{\rm dynamic}$, free variations lead to the vanishing of
$i_N\Delta\varphi$ and $i_N\Delta p$, which yields no formal
reduction of ${\cal B}_{\rm dynamic}$, while free variation for
${\cal H}_{\rm constrained}$ leads to $\Delta p=0=\Delta\varphi$,
consequently ${\cal B}_{\rm constrained}$ vanishes!  Hence we find
a curious fact: for free variations there is boundary expression
which is distinguished by having a vanishing value, it is the
expression which is complimentary to the one which vanishes for
trivial reference values.

Lanczos has discussed, via a detailed example, how one could do
more complicated things, where certain variables are
``controlled'' and others are varied ``freely''.  In all cases the
boundary variation principle yields the correct number of boundary
conditions.  For our expressions there are many possible
combinations of ``free''  and ``controlled'' variations.  Using
the ``natural boundary conditions'' associated with ``free
variation'' are an interesting option which merit further
exploration.  Based on our
 present understanding, however, for
most applications we favor purely ``controlled'' variations.

\section{The Phase space and asymptotics}

For finite regions (which is our primary interest) these boundary
terms in the variation of the Hamiltonian tell us exactly what
needs to be held fixed (not constant but rather ``controlled'',
i.e. the function on the boundary is predetermined by some outside
agent). For asymptotically flat regions, however, it is not
sufficient to just say we want the field to vanish at infinity.
Rather one should take into account the asymptotic fall off rates.
Asymptotically we want to allow the variations and responses to be
like the differences between generic solutions. The various
boundary terms we have constructed will all enable the Hamiltonian
to be well defined on the phase space of fields with asymptotic
behavior for all typical physical fields. Detailed investigations
have been done in particular for Einstein's gravity theory,
general relativity (GR) beginning with the pioneering work of
Regge and Teitelboim \cite{RT74}. This work was later improved by
Beig and \'O Murchadha \cite{BO87} and more recently further
refined by Szabados \cite{Sza03}.  Here we give a simplified
summary of certain relevant conclusions of these works.

It turns out that one should impose different fall off rates for
the terms with different parity.  For the fields it is sufficient
to take the respective asymptotic fall offs and parities to be
\begin{equation} \Delta\varphi\approx O^+_1+O^-_2,\qquad
\Delta p\approx O^-_2+O^+_3, \label{asymptotics}
\end{equation}
where $r^MO^\pm_M$ can be finite as $r\to\infty$ and $\pm$
indicates the parity. Here, for all types of tensors and forms, we
define the parity to mean the parity of the components in an
asymptotically Cartesian reference frame $dx^\nu$. Now the
Cartesian components of the 2-surface area element have odd
parity, consequently even parity 2-forms automatically have
vanishing 2-surface integral.

For asymptotically flat spaces the displacement should
asymptotically be a Killing vector, i.e. an infinitesimal
Poincar\'e displacement. More precisely one can allow
\begin{equation}
N^\mu \approx (\alpha_0 +\alpha^+_1)^\mu +
(\lambda_0+\lambda^+_1)^\mu{}_\nu x^\nu, \label{asykilling}
\end{equation}
where $\alpha_0^\mu$ is a constant translation parameter and
$\lambda_0^{\mu\nu}=\lambda_0^{[\mu\nu]}$ is a constant asymptotic
infinitesimal Lorentz boost/rotation parameter; the {\it even}
parity part of the perturbations in these parameters can have the
indicated $1/r$ fall off, but any {\it odd} parity perturbation to
$\lambda$ should fall off faster. Note that the $\lambda^+_1$
correction, because of its coordinate coefficient, means that it
can be regarded as an odd parity perturbation of $\alpha$.  Thus
we can, without any change in the conserved total values or
vanishing of the boundary term in the Hamiltonian variation, allow
{\it super-translations}---i.e.,
 asymptotically non-constant ($\equiv$ angular dependent) terms of
finite magnitude---as long as they are odd parity.  Even parity
supertranslations, on the other hand, would {\em change} the value
of the quasi-local quantities and would also yield in general a
{\em non-vanishing boundary term} in the Hamiltonian variation. If
one really wanted them one could allow even parity
supertranslations, but only at the expense of requiring boundary
conditions on the fields more strict than (\ref{asymptotics}).

With the asymptotics (\ref{asymptotics},\ref{asykilling}) it is
straightforward to verify, as we just stated, that all four of our
Hamiltonians are differentiable on the specified phase
space---since all the boundary terms in the variations of the
Hamiltonians vanish asymptotically; moreover the quasi-local
expressions all give {\it the same} (since they differ by terms of
the form $N\Delta\varphi\Delta p$ which vanish asymptotically)
{\it finite} constant values (independent of the perturbation
$\alpha^+_1$ and super-translation $\lambda^+_1x$) for the
energy-momentum and angular momentum/center-of-mass---even for
fields, like the metric or frame, with $\bar\varphi\approx1$.

Note that the indicated asymptotic behavior is sufficient but
hardly necessary. In practice only the specific projected
component combinations that actually show up in the boundary
integrand need be so restricted (moreover one need only require
this behavior up to a closed form). However it is not possible to
give a general {\it covariant} formula for such details; for such
refinements one must examine how each component for a particular
field in a specific theory actually occurs in the expression.

We also note that one could even turn things around and take the
finiteness of the quasi-local expressions and the vanishing of the
boundary term in the variation of the Hamiltonian as defining a
norm that would determine the largest possible phase space.  We do
not pursue this idea here.

At spatial infinity the aforementioned asymptotics are physically
reasonable. Let us now consider what can be expected if the
boundary of our 2-surface $\partial\Sigma$ approaches null
infinity.  One can imagine it following the characteristic
propagation surfaces of outgoing radiation. Long range radiation
fields (e.g. electromagnetism) have slower fall offs, like $\Delta
p \approx d\varphi\approx O_1$.  Then it may seem that we will
have a serious problem, since the boundary terms in the variation
of our various Hamiltonians will not vanish, so the Hamiltonian is
no longer functionally differentiable. This seeming calamity is
actually providential---as was recognized long ago (see e.g. the
remark on page 160 in \cite{Nes84}) but not investigated until
more recently---it is directing us to additional physics contained
within the formalism, namely energy flux expressions.

\section{Flux expressions}

Given any vector field $M=d/d\lambda$ one could calculate (on
shell) the associated change in any quasi-local quantity directly
from the boundary expression:
\begin{equation}
{d \over d\lambda}E_{\rm
B.C.}(N,\Sigma)=\int_\Sigma\!\!\Lie_M{\cal H}_{\rm
B.C.}(N)=\oint_{\partial \Sigma}\!\!\Lie_M{\cal B}_{\rm B.C.}(N).
\end{equation}
In this fashion one could calculate from such a ``flux''
expression, for example, the change in the quasi-local linear
momentum under a rotation or the time rate of change of the
angular momentum. In particular this approach can be specialized
to calculate the ``time rate of change'' of the ``energy''
associated with the ``time displacement'' $N=d/dt$ itself
\begin{equation}
{d\over dt}E_{\rm B.C.}(N,\Sigma)=\int_\Sigma \!\!\Lie_N{\cal
H}_{\rm B.C.}(N)= \oint_{\partial\Sigma}\!\! \Lie_N{\cal B}_{\rm
B.C.}(N). \label{dotB}
\end{equation}
The 2-surface integral then defines an ``energy flux''.  (Note: we
use the labels ``time'' and ``energy'' here for convenience in our
description, however the actual meaning is the change along $N$ in
the conserved quantity associated with $N$; e.g. we could consider
a rotation: $N=\partial/\partial\phi$).  Of course there is
another natural way to calculate the ``time rate of change'' of
any conserved quantity, namely one simply rearranges the conserved
current formula, such as $\partial_\mu j^\mu=0$ to $\partial_t
j^0=-\partial_k j^k$, and then integrates over a spatial region;
the spatial divergence via the divergence theorem yields a
boundary 2-surface flux integral. Such expressions are often
referred to as ``balance equations''. Expressed in terms of a
closed 3-form they follow just from integrating $\Lie_N j\equiv
di_N j$. In particular for energy this leads (on shell) to $\Lie_N
{\cal H}_{\rm B.C.}(N)\equiv di_N{\cal H}_{\rm B.C.}(N)\equiv di_N
d{\cal B}_{\rm B.C.}(N)\equiv d\Lie_N{\cal B}_{\rm B.C.}(N)$. Thus
this alternative approach also leads to (\ref{dotB}).

For the flux of ``energy'' (and {\it only} for the flux of {\it
energy}, i.e. the change of $E(N)$ along $N$, not for the ``time
rate of change'' of any other quantity like linear or angular
momentum, etc.) there is another way of calculating---the analog
of the classical mechanics calculation (for conservative
Hamiltonian systems) of
\begin{equation} \delta H=\dot q^k\delta p_k-\dot p_k\delta q^k
\qquad \Longrightarrow \qquad \dot E:=\dot H\equiv 0
\end{equation}
under the replacement $\delta\to d/dt$,
 where the remarkable
cancellation is a consequence of the particular (symplectic) form
of the Hamilton equations.
 Specializing the relations
(\ref{deltaHdir},\ref{deltaHneu},\ref{deltaHdyn},\ref{deltaHcon})
with $\delta\to\Lie_N$ (we are presuming that $\Lie_N
\bar\varphi$, $\Lie_N \bar p$ vanish, that could be a strong
restriction on the reference or on $N$), we note that the 3-form
parts vanish identically (again this is associated with the
symplectic form of the Hamilton equations), hence the respective
boundary flux expressions are the integrals of
\begin{eqnarray}
\!\!\!\!\!\Lie_N {\cal H}_{\rm Dirichlet}(N) &\!\!\!=\!\!\!& d i_N
(\Lie_N \varphi \wedge \Delta p),
\label{Dirflux}\\
\!\!\!\!\!\Lie_N {\cal H}_{\rm Neumann}(N) &\!\!\!=\!\!\!&  d i_N
(-\Delta \varphi \wedge \Lie_N p),
\label{Neuflux} \\
\!\!\!\!\!\Lie_N {\cal H}_{\rm dynamic}(N) &\!\!\!=\!\!\!& d
(\varsigma \Lie_N \varphi \! \wedge \! i_N \Delta p \! - \! i_N
\Delta \varphi \! \wedge \! \Lie_N p),
\label{dynflux}\\
\!\!\!\!\!\Lie_N {\cal H}_{\rm constraint}(N) &\!\!\!=\!\!\!& d
(i_N \Lie_N \varphi \! \wedge \! \Delta p \! - \! \varsigma \Delta
\varphi \! \wedge \! i_N \Lie_N p)\label{constrflux}.
\end{eqnarray}
These expressions hold even if ${\cal H}_\mu$ does not vanish
(i.e, when we have global but not local translation symmetry).
Note that they are significantly different in appearance from
those obtained using (\ref{dotB}) along with
(\ref{Bdir},\ref{Bneu},\ref{Bdyn},\ref{Bcon}):
\begin{eqnarray}
\!\!\!\!\Lie_N {\cal H}_{\rm Dirichlet}(N) &\!\!=\!\!& d \Lie_N
(i_N \varphi \wedge \Delta p -
\varsigma \Delta \varphi \wedge i_N \bar p), \label{Dirflux2}\\
\!\!\!\!\Lie_N {\cal H}_{\rm Neumann}(N) &\!\!=\!\!&  d \Lie_N
(i_N \bar \varphi \wedge \Delta p -
\varsigma \Delta \varphi \wedge i_N p),\label{Neuflux2} \\
\!\!\!\!\Lie_N {\cal H}_{\rm dynamic}(N) &\!\!=\!\!& d \Lie_N(i_N
\bar\varphi \wedge \Delta p -
\varsigma \Delta \varphi \wedge i_N \bar p), \label{dynflux2}\\
\!\!\!\!\Lie_N {\cal H}_{\rm constraint}(N) &\!\!=\!\!& d \Lie_N
(i_N \varphi \wedge \Delta p - \varsigma \Delta \varphi \wedge i_N
p),\label{constrflux2}
\end{eqnarray}
which hold only as long as ${\cal H}_\mu$ vanishes.  Via a
straightforward but non-trivial calculation using the field
equations (\ref{hameqs}) it can, of course, be explicitly verified
that the two respective forms are actually equivalent---when
${\cal H}_\mu$ vanishes.

\section{Application to Electromagnetism}
To illustrate these ideas first in a familiar setting let us
examine (vacuum) electromagnetism. One could consider this as a
source field for gravity, but it is more instructive, and still
sufficient for our needs here, to more simply just consider vacuum
electromagnetism in Minkowski space. In that case the formalism
developed above, with the important exception of the ``on shell''
vanishing of ${\cal H}_\mu$, is still applicable. The first order
Lagrangian 4-form for the (source free) U(1) gauge field one-form
$A$ is
\begin{equation}
{\cal L}_{\rm EM} = d A \wedge H - \frac1{2\lambda_0} \dual H
\wedge H,
\end{equation}
yielding the pair of first order equations
\begin{equation}
d H = 0, \qquad d A - \frac1{\lambda_0}\dual H = 0.
\end{equation}
These are just the vacuum Maxwell equations with $ \dual H =
\lambda_0F:=\lambda_0dA$; hence $H = - \lambda_0\dual F$, and $d
\dual F = 0$. (Here $\lambda_0^{-1}$ is the vacuum impedance.)
With $N=\partial_t$ and the decomposition $A=(-\phi, A_k)$ we find
that $i_N F=i_NdA=\Lie_NA-di_NA$ corresponds to $F_{0k}={\dot
A}{}_k +\partial_k \phi=-E_k$. The magnetic field strength is
$F_{ij}:=\partial_iA_j-\partial_jA_i =: \epsilon_{ijk} B^k$. Hence
$H_{0i}=-\lambda_0\dual F_{0i} = -\lambda_0B_i$,
$H_{ij}=-\lambda_0\dual F_{ij} = -\lambda_0\epsilon_{ijk} E^k$.
The natural reference choice is $\bar A = 0=\bar H$.

The Hamiltonian 3-form is
\begin{eqnarray}
{\cal H}^{\rm EM}_{\rm B.C.}(N) &=& - i_N A \, d H - d A \wedge
i_N H
\nonumber\\
&& + i_N \left( \frac1{2\lambda_0} \dual H \wedge H\right)+ d{\cal
B}_{\rm B.C.}.
\end{eqnarray}
In conventional tensor index notation the volume density part is
\begin{eqnarray}
{\cal H}^{\rm EM} &=& \lambda_0 \Bigg[ - \phi \partial_k E^k +
\frac12(\partial_i A_j - \partial_jA_i) F^{ij}
\nonumber\\
&& \qquad + \frac12 E^k E_k - \frac14 F^{ij} F_{ij} \Bigg].
\end{eqnarray}
After eliminating the 2nd class constraint,
$F_{ij}=\partial_iA_j-\partial_jA_i$, it corresponds to the
familiar $\lambda_0[\frac12(E^2+B^2)-\phi\partial_kE^k]$; the
scalar potential acts as a Lagrange multiplier to enforce the
Gauss constraint $\partial_kE^k=0$.

For the four considered boundary choices we have
\begin{eqnarray}
{\cal B}^{\rm EM}_{\rm Dir} &=& i_N A \, H = \lambda_0 \phi E^k
dS_k,
\\
{\cal B}^{\rm EM}_{\rm Neu} &=& A \wedge i_N H = - \lambda_0 A_i
B_j \epsilon^{ijk} dS_k,
\\
{\cal B}^{\rm EM}_{\rm dyn} &=& 0,
\label{EMdyn}\\
{\cal B}^{\rm EM}_{\rm con} &=& i_N A \, H + A \wedge i_N H
\nonumber \\
&=& \lambda_0 (\phi E^k - A_i B_j \epsilon^{ijk}) dS_k.
\end{eqnarray}
The Hamiltonian variations have the respective forms
\begin{eqnarray}
\delta{\cal H}^{\rm EM}_{\rm Dir}(N) &=& \hbox{field equation
terms} \nonumber\\
&&+ d(\delta i_N A\, H-\delta A\wedge i_N H),
\\
\delta{\cal H}^{\rm EM}_{\rm Neu}(N) &=&  \hbox{field equation
terms} \nonumber \\
&&+ d(-i_N A\, \delta H+A\wedge \delta i_N H),
\\
\delta{\cal H}^{\rm EM}_{\rm dyn}(N) &=&  \hbox{field equation
terms} \nonumber\\
&&+ d (  - i_N A \, \delta H- \delta A \wedge i_N H ),
\\
\delta{\cal H}^{\rm EM}_{\rm con}(N) &=&  \hbox{field equation
terms} \nonumber\\
&&+ d ( \delta i_N A \, H + A \wedge \delta i_N H).
\end{eqnarray}
Here our interest is not in the field equations but in the total
differential term which, upon integration, becomes  a boundary
term indicating the boundary condition. Briefly: the choice ${\cal
H}^{\rm EM}_{\rm Dir}$ corresponds to fixing the scalar potential
and the components of the vector potential parallel to the
2-surface (the gauge independent part of the latter fixes the
normal component $B_\bot$ of the magnetic field). ${\cal H}^{\rm
EM}_{\rm Neu}$ enables the fixing of certain projected components
of $F$, namely $ E_\bot$ and a part of $B_{||}$. The choice ${\cal
H}^{\rm EM}_{\rm dyn}$ is used for fixing $E_\bot$ and $B_\bot$,
while ${\cal H}^{\rm EM}_{\rm con}$ is used for fixing the scalar
potential and a part of $B_{||}$.

The physical meaning of such boundary conditions are well known
especially in the electrostatics case. Fixing $E_\perp$ on the
boundary corresponds to fixing the surface charge density. Here is
an instructive physical application: use a battery to first charge
up a capacitor which contains a dielectric which can be
inserted/removed, then disconnect the battery and measure the work
needed to remove/insert the dielectric (in the process the
potential will vary but the charge is fixed, no current or power
will flow into/out of the system, the system is {\it adiabatically
insulated}). Alternately leave the battery connected and measure
the work needed to displace the dielectric---then the potential is
fixed although the charge will now vary, so in this latter case
current and hence power flows into or out of the system.  The
respective boundary terms in the variation of the Hamiltonian are
$-\phi \delta E^k dS_k$ and $\delta\phi E^k dS_k$.  The point we
wish to emphasize is that both boundary condition choices are
physically meaningful; they correspond to real situations actually
encountered in practice.  Nevertheless there is a preferred
choice.

One expression stands out: the ${\cal H}^{\rm EM}_{\rm dyn}$
choice is the only one in which the value of the Hamiltonian is
{\it gauge invariant}. Moreover, in addition to the neat property
of enjoying a vanishing boundary term, it has an extra virtue of
considerable physical importance: namely the Hamiltonian density
is {\it non-negative}. Consequently the associated energy has a
lower bound and the system has a natural vacuum or ground state:
zero energy for vanishing fields.  The value of the Hamiltonian
$H^{\rm EM}_{\rm dyn}$ can be interpreted as the {\it internal
energy}, whereas the other expressions include some additional
energy on the boundary of the system associated with maintaining
the boundary condition.

The respective energy flux expressions from
(\ref{Dirflux}--\ref{constrflux}) are
\begin{eqnarray}
\!\!\!\!&\!\!\!\!& \Lie_N {\cal H}^{\rm EM}_{\rm Dir} \!=\! d
\left( i_N d i_N A \! \wedge \! H \! - \! d i_N A \! \wedge \! i_N
H \! - \! i_N F \! \wedge \! i_N H \right)
\nonumber\\
\!\!&\!\!&\quad\!\! = \lambda_0 d \left[(
\partial_i A_0 B_j \! - \! {\textstyle\frac12}
\partial_t A_0 \epsilon_{ijk} E^k \! - \! E_i B_j ) dx^i \!
\wedge \! dx^j \right]\!\!,
\\
\!\!\!\! &\!\!\!\!& \Lie_N {\cal H}^{\rm EM}_{\rm Neu}  =  d
\left( A \wedge i_N d i_N H - i_N A \wedge d i_N H \right)
\nonumber\\
\!\! &\!\!& \quad\!\!= \lambda_0 d \left[ \left( A_0
\partial_i B_j - A_i \partial_t B_j \right) dx^i \wedge dx^j
\right],
\\
\!\!\!\! &\!\!\!\!& \Lie_N {\cal H}^{\rm EM}_{\rm dyn} = - d
\left( i_N F \wedge i_N H \right)
\nonumber\\
\!\! &\!\!& \quad\!\!= \lambda_0 d \left(- E_i B_j \; dx^i \wedge
dx^j \right),
\\
\!\!\!\! &\!\!\!\!& \Lie_N {\cal H}^{\rm EM}_{\rm con} = d \left(
i_N d i_N A \wedge H + A \wedge i_N d i_N H \right)
\nonumber\\
\!\! &\!\!&\quad\!\! = \lambda_0d \left[ - \left(
{\textstyle\frac12}
\partial_t A_0 \epsilon_{ijk} E^k + A_i \partial_t B_j \right)
dx^i \wedge dx^j \right].
\end{eqnarray}
(Note we cannot calculate the correct energy flux in this case by
expressions like (\ref{dotB},\ref{Dirflux2}--\ref{constrflux2})
unless we include an additional non-vanishing contribution from
${\cal H}_\mu$.) All of these Hamiltonians and their associated
energies are really describing the same physical laws (the
differences in the right hand sides just correspond to differences
in the left hand sides), however all except ${\cal H}^{\rm
EM}_{\rm dyn}$ are boundary condition choices which are gauge
dependent descriptions. Clearly the ${\cal H}^{\rm EM}_{\rm dyn}$
choice, associated with fixing the normal components $E_\perp$ and
$B_\perp$ on the boundary, is preferred; it is the one suitable
for most physical applications. It gives the usual energy density
and Poynting flux.

\section{Application to Einstein's gravity theory}
Einstein's gravity theory, general relativity (GR), can be
formulated in several ways. For our purposes the most convenient
is to take the {\it orthonormal coframe}
$\vartheta^\mu=\vartheta^\mu{}_k dx^k$ and the {\it connection
one-form} coefficients
$\omega^\alpha{}_\beta=\Gamma^{\alpha}{}_{\beta k}dx^k$ as our
geometric potentials.  Moreover we take the connection to be {\it
a priori} metric compatible:
$Dg_{\alpha\beta}:=dg_{\alpha\beta}-\omega^\gamma{}_\alpha
g_{\gamma\beta}-\omega^\gamma{}_\beta g_{\alpha\gamma}\equiv0.$
Restricted to orthonormal frames where the metric coefficients are
constant, this condition reduces to the algebraic constraint
$\omega^{\alpha\beta}\equiv\omega^{[\alpha\beta]}$.

We consider the vacuum (source free) case for simplicity (the
inclusion of sources is straightforward).  GR can be obtained from
the first order Lagrangian 4-form
\begin{equation}
{\cal L}_{\rm GR} = \Omega^{\alpha\beta} \wedge \rho_{\alpha\beta}
+ D\vartheta^\mu\wedge\tau_\mu-V^{\alpha\beta} \wedge (
\rho_{\alpha\beta} - \frac1{2\kappa}\eta_{\alpha\beta}),
\end{equation}
where $ \Omega^\alpha{}_\beta := d \omega^\alpha{}_\beta +
\omega^\alpha{}_\gamma \wedge \omega^\gamma{}_\beta $ is the {\it
curvature} 2-form,
$D\vartheta^\mu:=d\vartheta^\mu+\omega^\mu{}_\nu\wedge\vartheta^\nu$
is the {\it torsion} 2-form,  and we have made use of the
convenient dual form basis $\eta^{\alpha\beta\dots}:=
\dual(\vartheta^\alpha \wedge \vartheta^\beta\wedge \cdots)$. The
2-forms $\Omega^{\alpha\beta}$, $V^{\alpha\beta}$ and
$\rho_{\alpha\beta}$ are antisymmetric. We take $\kappa:=8\pi
G/c^4=8\pi$.

In this Lagrangian the frame and connection conjugate momenta
$\tau_\mu$, $\rho_{\alpha\beta}$ and the auxiliary field
$V^{\alpha\beta}$ all function like Lagrange multiplier fields;
their variation imposes, respectively, the torsion vanishing
condition $D\vartheta^\mu=0$, the multiplier value
$V^{\alpha\beta}=\Omega^{\alpha\beta}$, and the conjugate momentum
value $\rho_{\alpha\beta}=(2\kappa)^{-1}\eta_{\alpha\beta}$. The
connection one-form variation gives (in vacuum)
\begin{equation}
D\rho_{\alpha\beta}+\vartheta_{[\beta}\wedge\tau_{\alpha]}=0.
\end{equation}
Since $D\rho_{\alpha\beta}\propto
D\eta_{\alpha\beta}=D\vartheta^\mu\wedge\eta_{\alpha\beta\mu}=0$
we get $\vartheta_{[\beta}\wedge\tau_{\alpha]}=0$, from which it
follows that $\tau_\mu=0$ (in vacuum). The frame variation gives
\begin{equation}
D\tau_\mu
+\frac1{2\kappa}V^{\alpha\beta}\wedge\eta_{\alpha\beta\mu}=0.
\end{equation}
Substituting the already determined values for $\tau_\mu$ and
$V^{\alpha\beta}$ yields
$\Omega^{\alpha\beta}\wedge\eta_{\alpha\beta\mu}=0$, the vanishing
of the Einstein tensor 3-form, i.e. the vacuum Einstein equation.

For most physical fields we can get by with a trivial reference.
However in the case of GR we certainly need the refinements of
both adjusting the Hamiltonian boundary term ``by hand''---as was
first clearly argued by Regge and Teitelboim (RT)
\cite{RT74}---and introducing a reference. With just the boundary
term in the Hamiltonian {\it naturally} inherited from the
Lagrangian, ${\cal
B}(N)=i_N\omega^{\alpha\beta}\wedge\rho_{\alpha\beta}$, the
boundary term in the variation of the Hamiltonian has the form
$i_N(\delta\omega^{\alpha\beta}\wedge\eta_{\alpha\beta})$, which
does not vanish for asymptotically flat fall offs.  A simple
reason for introducing the reference values, at least for the
connection, is to render covariant the manifestly non-covariant
``natural'' Hamiltonian boundary term.  (By the way the need for
reference values for GR was not apparent in the ADM \cite{ADM} and
RT formulations; there it was hidden in the choice of
asymptotically Cartesian coordinates.  The explicit need for a
reference metric in the asymptotic Hamiltonian boundary term was,
to our knowledge, first clearly apparent in the work of Beig and
\'O Murchadha \cite{BO87}).

In the vacuum case (or more generally as long as our boundary is
in the vacuum region) the frame conjugate momentum field
$\tau_\mu=0$, so only the connection and its conjugate momentum
make explicit contributions to our gravitational quasi-local
expressions. (When sources are included we can always choose the
boundary conditions so that the source fields have vanishing
quasi-local boundary term, however via the gravitational field
equations the sources, of course, indirectly influence the values
of the gravitational field variables on the boundary and thereby
do contribute to the quasi-local values.)

Now one of our dynamic gravitational variables is not a tensor
field, so there are certain terms in our quasi-local expressions
which include the non-covariant factors $i_N\omega^{\alpha\beta}$
or $i_N\bar\omega {}^{\alpha\beta}$. As discussed in more detail
in \cite{Chen:1998aw,Chen:2000xw}, the physical contribution due
to these terms is obtained by replacing them by $D^{[\beta}
N^{\alpha]}$ or ${\bar{D}}{}^{[\beta} N^{\alpha]}$.  Taking these
considerations into account, along with the identification
$\rho_{\alpha\beta}=(2\kappa)^{-1}\eta_{\alpha\beta}$, we obtain
our quasi-local GR Hamiltonian boundary term expressions for
gravitating systems (in vacuum regions):
\begin{eqnarray}
{\cal B}^{\rm GR}_{\vartheta}(N) &\!\!:=\!\!& \frac1{2\kappa}
\left( \Delta\omega^{\alpha\beta} \! \wedge \! i_N
\eta_{\alpha\beta} \! + \! {\bar{D}}{}^{[\beta} N^{\alpha]}
\Delta\eta_{\alpha\beta} \right),
\label{Btheta}\\
{\cal B}^{\rm GR}_{\omega}(N) &\!\!:=\!\!& \frac1{2\kappa} \left(
\Delta\omega^{\alpha\beta} \! \wedge \!
i_N{\bar\eta}_{\alpha\beta} \! + \! {{D}}{}^{[\beta} N^{\alpha]}
\Delta\eta_{\alpha\beta} \right),
\\
{\cal B}^{\rm GR}_{\rm dyn}(N) &\!\!:=\!\!& \frac1{2\kappa} \left(
\Delta\omega^{\alpha\beta} \!\wedge \! i_N{\bar\eta}_{\alpha\beta}
\! + \! {\bar{D}}{}^{[\beta} N^{\alpha]} \Delta\eta_{\alpha\beta}
\right),
\\
{\cal B}^{\rm GR}_{\rm con}(N) &\!\!:=\!\!& \frac1{2\kappa} \left(
\Delta\omega^{\alpha\beta} \! \wedge \! i_N\eta_{\alpha\beta} \! +
\! {{D}}{}^{[\beta} N^{\alpha]}\Delta\eta_{\alpha\beta}\right).
\end{eqnarray}
As already mentioned these quasi-local expressions will certainly
yield finite values when integrated over an asymptotic 2-sphere
for asymptotic Killing displacements of the form
(\ref{asykilling}),
at least when the variables satisfy the asymptotic parity and fall
off conditions
\begin{equation}\delta\vartheta\approx O^+_1+O^-_2,\qquad
\delta\omega\approx O^-_2+O^+_3.
\end{equation}
(At null infinity some parts of the connection actually have
slower fall off yet, as we shall see, the quasi-local integrals
are all still finite.)

The variations of the associated Hamiltonians have the respective
forms
\begin{eqnarray}
\delta{\cal H}^{\rm GR}_{\vartheta}(N) &=& \hbox{field equation
terms}
\\
&+& \frac1{2\kappa} d(- i_N \Delta \omega^{\alpha\beta} \, \delta
\eta_{\alpha\beta} + \Delta\omega^{\alpha\beta}\wedge\delta i_N
\eta_{\alpha\beta}),
\nonumber \\
\delta{\cal H}^{\rm GR}_{\omega}(N) &=& \hbox{field equation
terms}
\\
&+& \frac1{2\kappa} d (\delta i_N \omega^{\alpha\beta}
\,\Delta\eta_{\alpha\beta} - \delta\omega^{\alpha\beta} \wedge i_N
\Delta \eta_{\alpha\beta}),
\nonumber\\
\delta{\cal H}^{\rm GR}_{\rm dyn}(N) &=& \hbox{field equation
terms}
\\
&+& \frac1{2\kappa} d( - i_N \Delta \omega^{\alpha\beta} \,
\delta\eta_{\alpha\beta} - \delta \omega^{\alpha\beta} \wedge i_N
\Delta \eta_{\alpha\beta}),
\nonumber\\
\delta{\cal H}^{\rm GR}_{\rm con}(N) &=& \hbox{field equation
terms}
\\
&+& \frac1{2\kappa} d (\delta i_N \omega^{\alpha\beta} \, \Delta
\eta_{\alpha\beta} + \Delta \omega^{\alpha\beta} \wedge \delta i_N
\eta_{\alpha\beta}). \nonumber
\end{eqnarray}
Here our interest is not in the field equations but in the total
differential term which, upon integration, becomes  a boundary
term indicating the respective boundary conditions. Briefly:
${\cal H}^{\rm GR}_\vartheta$ requires fixing (after integration:
certain projected components of) the orthonormal coframe, while
${\cal H}^{\rm GR}_\omega$ requires fixing  (certain projected
components of) the connection. Whereas ${\cal H}^{\rm GR}_{\rm
dyn}$ is associated with fixing the spatial projections of the
frame and connection, and ${\cal H}^{\rm GR}_{\rm con}$ is
associated with fixing the time components of the frame and
connection. The boundary terms in these Hamiltonian variations
vanish asymptotically (spatially) with the aforementioned fall off
and parity conditions.  Asymptotically at spatial infinity the
quasi-local quantities obtained from all four of our expressions
are compatible with the analysis of Beig and \'O Murchadha
\cite{BO87}, or more precisely with the refinement thereof of
Szabados \cite{Sza03}.

The associated energy flux expressions, calculated according to
the respective prescriptions (\ref{Dirflux}--\ref{constrflux}),
(presuming that  $\Lie_N \bar\vartheta=0=\Lie_N\bar\omega$, i.e.
$N$ is a Killing field of the reference geometry) take the form
\begin{equation}
\Lie_N {\cal H}^{\rm GR}_{\vartheta}(N) \! = \! \frac1{2\kappa} d
( - i_N \Delta \omega^{\alpha\beta} \Lie_N \eta_{\alpha\beta} \! +
\! \Delta \omega^{\alpha\beta} \! \wedge \! \Lie_N i_N
\eta_{\alpha\beta}), \label{thetaflux}
\end{equation}
\begin{equation}
\Lie_N{\cal H}^{\rm GR}_{\omega}(N) \! = \! \frac1{2\kappa} d (
\Lie_N i_N \omega^{\alpha\beta} \Delta \eta_{\alpha\beta} \! - \!
\Lie_N \omega^{\alpha\beta} \! \wedge \! i_N \Delta
\eta_{\alpha\beta}), \label{omegaflux}
\end{equation}
\begin{equation}
\Lie_N{\cal H}^{\rm GR}_{\rm dyn}(N) \! = \! \frac1{2\kappa} d (
\! - \! i_N \Delta \omega^{\alpha\beta} \Lie_N \eta_{\alpha\beta}
\! - \! \Lie_N \omega^{\alpha\beta} \wedge i_N \Delta
\eta_{\alpha\beta}), \label{grdynflux}
\end{equation}
\begin{equation}
\Lie_N {\cal H}^{\rm GR}_{\rm con}(N) \! = \! \frac1{2\kappa} d (
\Lie_N i_N \omega^{\alpha\beta} \Delta \eta_{\alpha\beta} \! + \!
\Delta \omega^{\alpha\beta} \! \wedge \! \Lie_N i_N
\eta_{\alpha\beta}). \label{grconflux}
\end{equation}
Only one, $\Lie_N{\cal H}^{GR}_\vartheta$ (\ref{thetaflux}), is
free from non-covariant factors.

Of course one should check the actual values given by these
expressions. We have reexpressed (\ref{Btheta},\ref{thetaflux}) in
the asymptotically null regime using the NP spin coefficient
formalism and obtained good results \cite{WCN05}. Moreover we are
presently working on adapting our expressions to conformal
infinity; the results of that investigation will be reported in
due course. Meanwhile, similar to the calculations in
\cite{Hecht96}, we have tested all of these energy and energy flux
expressions at null infinity on the full Bondi-Sachs \cite{Sachs}
form of the metric using Reduce and EXCALC. In particular, for the
value of the expression ${\cal B}^{\rm GR}_\vartheta$ (which had
been calculated earlier \cite{Hecht96}) we obtain results
essentially identical to those reported in \cite{CJM98} obtained
using SHEEP to evaluate the Freud holonomic expression (which is
the expression Trautman used in his original work
\cite{Trautman}). This is not at all surprising, since this
particular orthonormal frame expression, for any frame satisfying
the asymptotic ``no rotation'' gauge condition $\vartheta_{[\alpha
k]}\approx O_2$, is asymptotically equivalent to both the Freud
superpotential \cite{Chen:1998aw,Chen:2000xw} and the expression
of Katz and coworkers \cite{Katz97}; the latter also gives good
results at null infinity.

Our three other boundary expressions give similar but not
identical results. Both for brevity and to more clearly show the
main similarities and differences we here present only the
essential details for the simpler, axi-symmetric Bondi metric
\cite{Bondi}. We take the coframe to be of the form
\begin{eqnarray}
\vartheta^t = e^{\beta+\phi}du + e^{\beta-\phi}dr, &\quad&
\vartheta^r = e^{\beta-\phi}dr, \nonumber\\
\vartheta^\theta = r e^\gamma (d\theta-r^{-2}U du), &\quad&
\vartheta^\varphi = r e^{-\gamma} \sin\theta d\varphi,
\end{eqnarray}
where $e^{2\phi} = 1 - 2m(u,\theta)/r + O_2$, $\gamma =
c(u,\theta)/r + O_2$, $\beta = - c^2/4 r^2 + O_4$, and $U = -
\partial_\theta c - 2 c \cot\theta + O_1$. For the reference we take
$\phi = \gamma = \beta = U = 0$. For $N = \partial_u$ we find for
the asymptotically contributing part of the energy expressions
\begin{equation}
{\cal B}^{\rm GR}_\vartheta = {\cal B}^{\rm GR}_{\rm con} =
\frac1{2\kappa} [4 m \sin\theta + 2\partial_\theta (U\sin\theta )]
\, d\theta \wedge d\varphi, \label{BondiE}
\end{equation}
\begin{equation}
{\cal B}^{\rm GR}_\omega = {\cal B}^{\rm GR}_{\rm dyn} =
\frac1{2\kappa} [4(m \!+ \! cc_u)\sin\theta \!\! + \!\!
2\partial_\theta (U \! \sin\theta )] d\theta \! \wedge \!
d\varphi.\label{notBondiE}
\end{equation}
Upon integration over the sphere the terms with
$\partial_\theta(U\sin\theta)$  make a vanishing contribution due
to the regularity conditions on $c$, $\partial_\theta c$ at the
poles (note: similar terms appear in the 2-surface integrands in
\cite{Hecht96,CJM98}). Then the two expressions in (\ref{BondiE})
give the standard Bondi mass aspect, while the other two do not.

The asymptotically contributing part of the associated quasi-local
flux values were found to be
\begin{eqnarray}
\Lie_{\partial_u} {\cal H}^{\rm GR}_\vartheta &\!\!=\!\!&
\Lie_{\partial_u} {\cal H}^{\rm GR}_{\rm con}
=\frac1{2\kappa}d\left[-4c_u^2 \sin\theta\, d\theta \wedge
d\varphi\right] ,\label{Bondiflux}
\\
\Lie_{\partial_u} {\cal H}^{\rm GR}_\omega &\!\!=\!\!&
\Lie_{\partial_u} {\cal H}^{\rm GR}_{\rm dyn}
=\frac1{2\kappa}d\left[4 cc_{uu}\,\sin\theta\, d\theta \wedge
d\varphi\right] .\label{notBondiflux}
\end{eqnarray}
Thus two expressions (\ref{Bondiflux}) directly give the standard
Bondi flux loss.  The other two expressions are actually
describing exactly the same physics albeit for a different energy
expression, as can be easily seen from comparing the time
derivative of the expression (\ref{BondiE}) equated to
(\ref{Bondiflux}) with the time derivative of the expression
(\ref{notBondiE}) equated to (\ref{notBondiflux}). Note that one
can equally well compute an incoming flux from past null infinity
by assuming a dependence on the advanced time $v\simeq t+r$ in
place of the retarded time $u\simeq t-r$.

 From these calculations we can see that the functions $\beta$ and
$U$ do not play a major role in this limit.  This observation
justifies the following simpler calculation, which still includes
enough of the essential qualities while  showing that our
expressions capture both the incoming and outgoing quasi-local
flux.   Let us evaluate the energy and energy flux expressions for
the following simplified asymptotic form of the orthonormal
coframe:
\begin{eqnarray}
\vartheta^t &=& e^\phi dt,
\nonumber\\
\vartheta^r &=& e^{-\phi} dr,
\nonumber\\
\vartheta^\theta &=& e^\gamma r d\theta,
\nonumber\\
\vartheta^\varphi &=& e^{-\gamma} r \sin\theta d\varphi,
\end{eqnarray}
where $\phi=\phi(t,r)=O(1/r)$, $\gamma=\gamma(t,r)=O(1/r)$.  For
the reference we take $\phi=0=\gamma$.  The connection one-form
components are
\begin{eqnarray}
\omega^{tr} &=& \phi' e^{2\phi} dt - \dot \phi e^{-2\phi} dr,
\nonumber\\
\omega^{t\theta} &=& r \dot\gamma e^{-\phi + \gamma} d\theta,
\nonumber\\
\omega^{t\varphi} &=& - r \dot\gamma e^{-\phi-\gamma} \sin\theta\,
d\varphi,
\nonumber\\
\omega^{r\theta} &=& -(1+r\gamma') e^{-\phi+\gamma} d\theta,
\nonumber\\
\omega^{r\varphi} &=& - ( 1 -r\gamma') e^{-\phi-\gamma}
\sin\theta\, d\varphi,
\nonumber\\
\omega^{\theta\varphi} &=& - e^{-2\gamma} \cos\theta\, d\varphi.
\end{eqnarray}
For our expressions with $N=\partial_t$ the $DN$ terms make no
asymptotic contribution.  The key factors are
\begin{eqnarray}
\Delta\omega^{r\theta} &=& \left[1 -
(1+r\gamma')e^{\phi+\gamma}\right]\,d\theta,
\nonumber\\
\Delta\omega^{r\varphi} &=& \left[1-
(1-r\gamma')e^{\phi-\gamma}\right]\, \sin\theta\, d\varphi.
\end{eqnarray}
 From these we find the asymptotically contributing part of the
quasi-local energy boundary expressions:
\begin{eqnarray}
{\cal B}^{\rm GR}_\vartheta &=& {\cal B}^{\rm GR}_{\rm con}=
\frac1{2\kappa}4 m \sin\theta d\theta \wedge d\varphi,
\label{bondi}\\
{\cal B}^{\rm GR}_\omega &=& {\cal B}^{\rm GR}_{\rm dyn}=
\frac1{2\kappa}4[m-r^2\gamma\gamma'] \sin\theta\, d\theta \wedge
d\varphi
\nonumber\\
&=& \frac1{2\kappa}4[m-r^2\gamma(\gamma_v-\gamma_u)] \sin\theta\,
d\theta \wedge d\varphi,\label{notbondi}
\end{eqnarray}
where, as usual, $m:=(r/2)(1-e^{2\phi})$ and $u:=t-r$, $v:=t+r$.
Upon integration over the 2-sphere $r={\rm const}$ we find that
two boundary conditions, corresponding to the boundary terms
${\cal B}^{\rm GR}_\theta$ and ${\cal B}^{\rm GR}_{\rm con}$, give
the Bondi mass.

The associated quasi-local flux values are found to be
\begin{eqnarray}
\Lie_N {\cal H}^{\rm GR}_\vartheta &\!\!=\!\!& \Lie_N {\cal
H}^{\rm GR}_{\rm con} = \frac1{2\kappa} d \left[ 4 r^2 \dot\gamma
\gamma' \sin\theta\, d\theta \wedge d\varphi \right]
\nonumber\\
&\!\!=\!\!& \frac1{2\kappa}d\left[4r^2(\gamma_v^2-\gamma_u^2)
\sin\theta\, d\theta \wedge d\varphi\right], \label{bondiflux}
\\
\Lie_N {\cal H}^{\rm GR}_\omega &\!\!=\!\!& \Lie_N {\cal H}^{\rm
GR}_{\rm dyn} = \frac1{2\kappa} d\left[ -4 r^2 \gamma
\dot\gamma{}' \sin\theta\, d\theta \wedge d\varphi \right]
\nonumber\\
&\!\!=\!\!& \frac1{2\kappa} d \left[ - 4 r^2 \gamma (\partial^2_v
\gamma - \partial^2_u \gamma) \sin\theta\, d\theta \wedge d\varphi
\right]. \label{notbondiflux}
\end{eqnarray}
Integrating  (\ref{bondi}--\ref{notbondiflux}) over a large
2-sphere at constant $r$, we find that all control modes are
consistent with a Bondi news type energy flux loss and gain from
both outgoing and incoming radiation:
\begin{equation}
\dot m = r^2\dot\gamma\gamma'=r^2(\gamma_v^2-\gamma_u^2).
\end{equation}
However only the Hamiltonians ${\cal H}^{\rm GR}_\theta$ and
${\cal H}^{\rm GR}_{\rm con}$ give this relation {\it directly}.
For most applications ${\cal H}^{\rm GR}_\theta$ would be the
preferred choice, because the associated flux relation
(\ref{thetaflux}) is free of non-covariant factors.


\section{Discussion}

 The Hamiltonian for a gravitating system in a finite or infinite
 region necessarily includes a boundary term.  We have inquired
 into the significance and best form of this boundary term.  We
 found that it  determines not only the quasi-local values but
 also, via  the {\it boundary variation principle}, the boundary
 conditions necessary for a well-defined Hamiltonian.  We noted
 that it is always possible (and necessary, at least for gravity)
 to include in it non-dynamic reference values for the dynamic
 variables. Using our covariant Hamiltonian formalism and an
 identity associated with the variation of the Hamiltonian, we have
 identified, for each dynamic field, four special quasi-local
 energy-momentum boundary term expressions; each corresponds to a
 physically distinct and geometrically clear boundary condition. We
 showed how a fundamental Hamiltonian variation identity naturally
 forces us, for radiating asymptotics, to relax the well-defined
 Hamiltonian requirement and thereby obtained the associated
 quasi-local energy flux expressions.  When the formalism is
 applied to electromagnetism one of the four is distinguished by
 gauge invariance (it gives the familiar energy density and
 Poynting flux). When the formalism is applied to Einstein's
 general relativity, two different boundary condition choices
 correspond to quasi-local expressions which asymptotically give
 the ADM energy, the Trautman-Bondi energy and, moreover, an
 associated energy flux (both outgoing and incoming), but once
 again there is a distinguished expression: in this case it is the
 one which is covariant.

 Here we make a few further remarks. First we mention that our
 ideas regarding variational principles, symplectic structure and
 the role of boundary terms have been much influenced by several
 sources, especially Tulczyjew and Kijowski \cite{KT79,Kij97} and
 his coworkers. Although we considered radiation using the
 Hamiltonian, there are many aspects associated with the
 Hamiltonian in the radiating regime that we did not discuss; these
 issues are nicely treated in \cite{CJK02}. The uniqueness of the
 Trautman-Bondi mass is established in \cite{CJM98}.

 We, like many others, took a Hamiltonian approach. While some of
 our expressions are similar to those that appear in other
 approaches, e.g.\ the Noether charge approach, our formalism
 includes certain features that are unusual. Features which
 distinguish our formalism include:
  We have made
 extensive use of differential forms---because of their convenience
 for integration over domains with boundaries, the space-time
 decomposition, as well as the representation of geometric and
 gauge fields. We started from a first order Lagrangian---mainly
 because it facilitates the passage from the Lagrangian to the
 Hamiltonian but also because of its linearity re coupling via
 connections. We used the (co)frame and connection as independent
 variables. The  Hamiltonian formalism we have developed is
 4-covariant.

 Our aim has been to find 4-D covariant 2-form and 3-form
 expressions which can be integrated over any suitable desired
 regions; essentially we want to find a good Hamiltonian density
 for the region of interest. In this approach the particular
 3-surface $\Sigma$ and its boundary $\partial\Sigma$ are quite
 incidental. Accordingly, our reference values and displacement are
 independent of these surfaces.   Note also that we do not
 decompose our expressions into intrinsic and extrinsic parts with
 respect to such surfaces, as was done both traditionally and in
 many modern works, e.g. \cite{AT02a,AT02b} wherein some of the
 themes considered here have been treated with the aid of such
 decompositions.

 Aside from some important clarifications of the formalism, the
 main new ingredients here involve our variational identity, the
 relaxation of the boundary variational principle, and the flux
 expressions associated with our four
 quasi-local expressions, along with the fact that in each
 application one expression is distinguished (by being {\it gauge
 invariant} for the relevant type of gauge covariance---$U(1)$ or
 Lorentz in our electromagnetic and gravitational examples,
 respectively).
  Although in our discussion we referred  to {\it energy} and {\it
 energy-flux},
 what we really  considered is the value of the Hamiltonian for a
 prescribed displacement $N$; the associated flux expression that
 gives the rate of change of this value along $N$.  Our formal
 results may have other useful applications besides energy.

 Localization of energy-momentum has been an outstanding problem
 from the very beginning of GR. Traditional methods, both Noether
 spacetime translation symmetry and decomposition of the field
 equations, have only led to a variety of reference frame dependent
 expressions (generally referred to as pseudotensors) for the
 energy-momentum density.  Thus in addition to the ambiguity of
 which expression should be preferred, there was also the ambiguity
 of the choice of reference frame. The boundary variation principle
 along with the Hamiltonian formalism tamed these inherent
 ambiguities, giving them a clear physical and geometric
 interpretation: the choice of expression is related to the choice
 of boundary conditions and the choice of reference frame is
 associated with the choice of ground state
 \cite{Chang:1998wj,Chen:2000xw,Nes04}. Within the covariant
 Hamiltonian formalism using this principle we had identified
 certain special ``covariant symplectic'' Hamiltonian boundary term
 quasi-local expressions \cite{Chen:1994qg,Chen:1998aw,Chen:2000xw}.
  Here, based on a fundamental Hamiltonian
 identity, we have identified the associated {\it energy flux}
 expressions.  Moreover we found that, for both source fields and
 gravity, among the four {\it one} particular quasi-local
 Hamiltonian boundary term and its associated flux expression was
 distinguished.

 An additional virtue enjoyed by the respective distinguished
 boundary terms (\ref{EMdyn},\ref{Btheta}) is that the Hamiltonian
 for this quasi-local choice has {\it positive energy}, not only
 for electromagnetism (as we discussed above), but also for
 gravity. More precisely (for acceptable choices of the variables)
 the gravitational Hamiltonian with the distinguished boundary term
 (\ref{Btheta}) is non-negative, at least when $\Sigma$ is maximal
 or nearly maximal. Here we just make mention of two lines of
 argument that lead to this conclusion. One can obtain this result
 by adapting to finite regions the global positivity proof
 \cite{Nes89} using the SOF gauge conditions \cite{SOF}; one can
 instead use an adaptation of the Shi and Tam proof \cite{ST02},
 which guarantees the positivity (for mean convex 2-surfaces) of the Brown-York \cite{BY93}
 quasi-local expression (the latter, as shown in
 \cite{Chen:1998aw}, agrees with (\ref{Btheta}) for certain
 choices). We are presently working on a detailed account of these
 two arguments.  Of course it would be nice to have a stronger and
 more general energy-momentum positivity proof for our
 distinguished covariant-symplectic quasi-local GR Hamiltonian
 boundary term.

\section*{Acknowledgments}
\medskip

The work of C.~M.~C. and J.~M.~N.  was supported by grants from
the National Science Council of the Republic of China under the
grant numbers NSC 93-2112-M-008-021,  NSC 93-2112-M-008-001 and
NSC 94-2119-M-002-001, while the work of R.~S.~T.  was supported
by the National Nature Science Foundation of China under the grant
numbers 10375081 and 10375087. J.~M.~N. appreciates his
discussions with L\'aszl\'o Szabados which helped to clarify some
of these issues.


\end{document}